\documentstyle[12pt]{article}

\newcommand{\be}{\begin{equation}}
\newcommand{\ee}{\end{equation}}
\newcommand{\bear}{\begin{eqnarray}}
\newcommand{\eear}{\end{eqnarray}}
\newcommand{\dr}{{d R(t)\over dt}}
\newcommand{\di}{{d I(t)\over dt}}
\newcommand{\dtp}{{d P(t)\over dt}}
\newcommand{\dI}{{d {\cal I}(t)\over dt}}
\newcommand{\dR}{{d {\cal R}(t)\over dt}}
\newcommand{\dP}{{d {\cal P}(t)\over dt}}
\newcommand{\dPP}{{d^2 {\cal P}(t)\over dt^2}}
\newcommand{\dPPP}{{d^3 {\cal P}(t)\over dt^3}}
\newcommand{\ga}{\Gamma_{\perp}}
\newcommand{\gaa}{\Gamma_{\parallel}}
\newcommand{\cube}{ \left (-{q\over 2} + \sqrt{\left({p\over 3}\right)^3 + 
\left({q\over 2}\right)^2} 
\right)^{1\over 3}}
\newcommand{\pbth}{ \left ({p\over 3}\right) }
\begin{document}
\title{Neutrino propagation in a random magnetic field}
{\author{Sarira Sahu$^{1}$\\
Theory Group, Physical Research Laboratory,\\
Navrangpura, Ahmedabad-380 009\\ India}
\date{ }
\maketitle
\footnotetext[1]{email:sarira@prl.ernet.in}
\thispagestyle{empty}
\begin{abstract}

\noindent The active-sterile neutrino conversion probability is calculated
for neutrino propagating in a medium in the presence of random
magnetic field fluctuations. Necessary condition for the probability
to be positive definite is obtained. Using this necessary condition 
we put constraint on the neutrino magnetic moment from active-sterile
electron neutrino conversion in the early universe hot plasma and
in supernova.
\end{abstract}
\vfill
\eject

\section{Introduction}

Neutrino propagation in the presence of a strong magnetic field is of
great interest from the point of view of astrophysics as well as 
early universe hot plasma\cite{zeldo}.  
The presence of primordial magnetic fields ($B\simeq 10^{-6}$G) 
over galactic scales, when extrapolated back can give very large fields and
these large fields might have affected the particle
interactions in the early universe\cite{kron}. 
It is believed that the primordial plasma was chaotic in nature
and the magnetic flux lines moving along with this plasma might get mixed up
creating randomness in the fields itself. It is also believed that 
the magnetic field inside a newly born neutron star is huge and random
in nature\cite{raffelt,raffelt1,thomson,ruder}. 
This strong random magnetic fields can influence 
the neutrino propagation as well as their conversion in the early 
universe
hot plasma and in a supernova medium\cite{kron,athar,pas1,pas2}.
Neutrinos having magnetic moment or transition magnetic 
moment can flip their chirality and left handed neutrinos becomes right handed
sterile neutrinos in the presence of an external magnetic field. These right
handed neutrinos being sterile to weak interaction stream out from the sun
or supernova core. The mechanism of helicity flipping in the presence of
a magnetic field  is one of
the explanation for the supernova cooling mechanism or the solar neutrino 
deficit\cite{vol}. Recently Pastor et. al.\cite{pas1}
have derived bounds on the transition 
magnetic moment of the Majorana neutrinos in the presence of  random
magnetic fields from the supernova energy loss as well as from nucleosynthesis.
In this paper we discuss about the active-sterile electron neutrino
conversion ($\nu_{eL}\rightarrow\nu_{eR}$) in the presence of a random
magnetic field in early universe hot plasma and in supernova medium. 
From the positive definiteness of the average conversion probability
we put constraint on the neutrino magnetic moment.

The paper is organized as follows: In Section 2, 
we have derived the average probability equation for the
active sterile/active $(\nu_a\rightarrow\nu_x)$ neutrino conversion 
in the presence of a random fluctuation over the
constant background magnetic field. For this derivation 
we have used the simple delta correlation for the uncorrelated random
magnetic field domains. We show that, the fluctuation in the transverse
and longitudinal components of the magnetic field are mixed up.
We obtain the solution for the probability equation and found  the
necessary condition for the conversion probability to be positive
definite. We have considered the neutrino propagation in the early universe
plasma and in the supernova medium. 
Assuming the magnetic fields in the early universe hot plasma and inside the
core of a newly born neutron star to be strong and purely random in nature
we put constraints on the magnetic moment for active sterile neutrino
conversion, which are discussed in Section 3 and 4 respectively. 
In discussion Section 5, we briefly summarise our results and discuss 
about the further improvement of the bound on the neutrino magnetic moment.

\section{Neutrino propagation}

The wave equation for the propagation of two neutrinos, 
one active and one
sterile (or active) in a plasma in the presence of a magnetic field is
given by
\be
i{d\over dt}{\pmatrix {\nu_a\cr \nu_x\cr}}
 = {\pmatrix {H_{aa} & H_{ax}\cr H_{xa} & H_{xx}\cr}}
 {\pmatrix {\nu_a\cr \nu_x\cr}}
\label{evu}
\ee
 where $x=s,b$ (sterile, active) and $b = e,\mu$ and $\tau$.
In the evolution eqn.(\ref{evu}) the diagonal components are
$H_{aa}=V_{vec}-\Delta + 
V_{axial}$ and $H_{xx}=0$ and
the off-diagonal entries   
$H_{ax}=H_{xa} = \mu B_{\perp}(t)$.
The $\Delta$ is given as
$\Delta=\cos(2\theta)~ (m_2^2-m_1^2)/2E$ and vanishes for 
degenerate Majorana neutrino and E is the neutrino energy.
For Dirac neutrino $\Delta = 0$ and
$\mu$ corresponds to its diagonal magnetic moment. On the other hand for
Majorana neutrino $\mu$ is the transition magnetic moment.
$V_{vec}$ is the difference of neutrino
vector interaction potentials, $V_{vec} = V^{\nu_a}_{vec} - V^{\nu_x}_{vec}$.
Here we consider neutrinos without mixing, which implies $\cos(2\theta)=1$.
The axial vector potential $V_{axial} = \mu_{eff}{{\bf k.B}/ k}$
is generated by the mean axial vector
current of charged leptons in an external 
magnetic field \cite{raffelt1,valle,espo}.

Let us define the functions $R=Re (<\nu_a^*\nu_x>)$ and 
$I=Im (<\nu_a^*\nu_x>)$. Then using these in eqn.(\ref{evu})
 we obtain the following set of equations:
\be
\dtp = -2 H_{ax}(t) I(t),
\label{pt}
\ee
\be
\di = H_{aa}(t) R(t) + H_{ax}(t) (2 P(t) -1),
\label{it}
\ee 
and
\be
\dr = -H_{aa}(t) I(t).
\label{ir}
\ee
The function $P(t)$ is the neutrino conversion probability 
$P_{\nu_a\rightarrow \nu_x}(t)$.
The magnetic field can be written as
$B(t)=B_0+{\tilde B}(t)$, where $B_0$ is the  constant background
field and ${\tilde B}(t)$ is the random fluctuation over it.
Putting this in $H_{aa}(t)$ and $H_{ax}(t)$ we get
\be
H_{aa}(t) = H_{aa}(0)+{\tilde H_{aa}}(t) 
= (V_{vec}-\Delta + \mu_{eff}B_{\parallel 0})
+\mu_{eff}{\tilde B_{\parallel}}(t)
\label{haa}
\ee
and
\be
H_{ax}(t) = H_{ax}(0)+{\tilde H_{ax}}(t) 
          = \mu B_{\perp 0}+ \mu{\tilde B_{\perp}}(t).
\label{hax}
\ee
For the neutrino conversion length 
greater than the domain size ($l_{conv}~>>~L_0$) one can average the 
the equations(\ref{pt}), (\ref{it}) and (\ref{ir}) over the random
magnetic field distribution. Let us define the average of the
functions $<P(t)>~=~{\cal P}(t)$,
$<R(t)>~=~{\cal R}(t)$ and
$<I(t)>~=~{\cal I}(t)$. Using these in eqns.(\ref{it}) and (\ref{ir}) we
obtain
\bear
\dI &=& <H_{ax}(t) (2P(t)-1)> 
+ <H_{aa}(t)R(t)>\nonumber\\
&=& H_{ax}(0) <(2P(t)-1)> + <{\tilde H_{ax}}(2P(t)-1)>
\nonumber\\
& & +H_{aa}(0)<R(t)> + <{\tilde H_{aa}}(t)R(t)>,
\eear
and
\be
\dR = -H_{aa}(0)<I(t)> - <{\tilde H_{aa}}(t) I(t)>
\ee
respectively.
The magnetic field in different domains is randomly oriented with respect to
the neutrino propagation direction. So the neutrino conversion probability
depends on the root mean square ({\it rms}) value of the random magnetic field.
With the use of the delta correlation 
for uncorrelated magnetic field domain of size $L_0$, the average of the
random magnetic field is\cite{semikoz,enq1,enq2},
\be
<B_{\parallel}(t)> =<B_{\perp}(t)> =<B_{\parallel}(t)B_{\perp}(t)>=0,
\ee
\be
<B_{i\parallel}(t)B_{j\parallel}(t_1)> 
=<B^2_{\parallel}>\delta_{ij}L_0\delta(t-t_1),
\ee
and
\be
<B_{i\perp}(t)B_{j\perp}(t_1)> 
=<B^2_{\perp}>\delta_{ij}L_0\delta(t-t_1).
\ee
The {\it rms} value of the averaged magnetic field is given as
$B_{rms} = \sqrt{<B^2>}$.
Neglecting the higher order correlation of $H_{aa}(t)$ and $H_{ax}(t)$
with $I(t)$, $R(t)$ and $P(t)$ and
using the above equations in eqn.(\ref{pt}) to (\ref{ir}) we obtain
\be
\dI =
H_{aa}(0) {\cal R}(t) + H_{ax}(0) (2 {\cal P}(t) - 1)
-2 (\ga + \gaa) {\cal I}(t),
\ee
\be
\dR = - H_{aa}(0) {\cal I}(t) - 2 \gaa {\cal R}(t),
\ee
and
\be
\dP = -2H_{ax}(0){\cal I}(t)-{\ga}\left (2{\cal P}(t)-1
\right ).
\label{pdot}
\ee
For convenience we define
\be
{\cal I}(t) = e^{-2(\Gamma_{\parallel}+\Gamma_{\perp})t}{\cal I}_{1}(t),
\ee     
and
\be
{\cal R}(t) = e^{-2\Gamma_{\parallel}t}{\cal R}_{1}(t),
\ee
where 
\be
\Gamma_{\perp}={4\over 3}\mu^2<B^2>L_0,
\ee
and
\be
\Gamma_{\parallel}={1\over 6}\mu_{eff}^2<B^2>L_0.
\ee
$\ga$ and $\gaa$ are the transverse and longitudinal magnetic field damping
parameters. 
Putting these in eqn.(\ref{pdot}) and differentiating two times with respect to 
$t$ we obtain
\bear
\dPPP &+&
4\left ( \Gamma_{\perp}+\Gamma_{\parallel}\right ) \dPP
+ 4\left ( 3\Gamma_{\perp}\Gamma_{\parallel}+
\Gamma_{\perp}^2+\Gamma_{\parallel}^2
+H^2_{ax}(0)+{H^2_{aa}(0)\over 4}\right )\dP
\nonumber\\
&+&8\left (\Gamma_{\perp}^2\Gamma_{\parallel}
+\Gamma_{\perp}\Gamma_{\parallel}^2
+H_{ax}^2(0)\Gamma_{\parallel}
+{H_{aa}^2(0)\Gamma_{\perp}\over 4}\right ){\cal P}(t)
\nonumber\\
&-&4\left (\Gamma_{\perp}^2\Gamma_{\parallel}
+\Gamma_{\perp}\Gamma_{\parallel}^2
+H_{ax}^2(0)\Gamma_{\parallel}+{H_{aa}^2(0)\Gamma_{\perp}\over4}\right )=0,
\label{master}
\eear
with the boundary conditions
${\cal P}(t)/_{t=0}=0,~~\dP/_{t=0}=\Gamma_{\perp}$ and 
$\dPP/_{t=0}=2H_{ax}^2(0)-2\Gamma_{\perp}^2$.
Switching off the damping terms in eqn.(\ref{master}), will give
MSW type solution  with magnetic field\cite{athar,vol,athar1}.
For $H_{aa}(0)=H_{ax}(0)=\gaa=0$, the eqn.(\ref{master})
reduces to 
\be
\dPP + 4\ga \dP + 4 \ga^2 {\cal P}(t) - 2 \ga^2 =0,
\ee
and it has the same solution as shown by Pastor et. al.\cite{pas1}. 
In this the effect of strong random magnetic field on
the neutrino transition magnetic moment is studied both in the
early universe hot plasma and in supernova.
The eqn.(\ref{master}) can be written in a simplified form as
\be
\dPPP+ A_0 \dPP+B_0 \dP +2 C_0 {\cal P}(t) - C_0 = 0,
\label{master2}
\ee
where the quantities $A_0$, $B_0$ and $C_0$ are
\be
A_0=4\left ( \Gamma_{\perp}+\Gamma_{\parallel}\right ),
\ee
\be
B_0= 4\left ( 3\Gamma_{\perp}\Gamma_{\parallel}+
\Gamma_{\perp}^2+\Gamma_{\parallel}^2
+H^2_{ax}(0)+{H^2_{aa}(0)\over 4}\right ),
\label{bb}
\ee
and 
\be
C_0=4\left (\Gamma_{\perp}^2\Gamma_{\parallel}
+\Gamma_{\perp}\Gamma_{\parallel}^2
+H_{ax}^2(0)\Gamma_{\parallel}+{H_{aa}^2(0)\Gamma_{\perp}\over4}\right )
\label{cc}
\ee
respectively. 
Eqns.(\ref{bb}) and (\ref{cc}) shows that the damping terms are mixed up
among themselves and also with the terms $H_{aa}(0)$ and $H_{ax}(0)$.
The solution of eqn.(\ref{master2}) is 
\be
{\cal P}(t) = y(t) + {1\over 2},
\label{sol}
\ee
where $y(t)$ is given as
\bear
y(t)& =& e^{-({{Z_1\over 2} + {A_0\over 3})t}}
\left [A_1
\left \{ e^{{3\over 2} Z_1 t} - \cos(Z_4 t) - 
{3\over 2}{Z_1\over Z_4} \sin(Z_4 t)\right \}  \right.
\nonumber\\
&& \left. -{\cos(Z_4 t)\over 2} + 
{(\Gamma_{\perp} - {A_0\over 6}-{Z_1\over 4})\over Z_4} \sin(Z_4 t)\right ].
\label{yt}
\eear
The coefficient $A_1$ is given as 
\be
A_1=-{(A^2_0 - 12 A_0\Gamma_{\perp} + 36\Gamma_{\perp}^2 - 36 H^2_{ax}(0)
+ 3 A_0 Z_1 - 18\Gamma_{\perp} Z_1 + {9\over 4} Z_1^2 + 9 Z_4^2)\over
({81\over 2} Z_1^2 + 18 Z_4^2)}, 
\ee
and $Z_1$ and $Z_4$ are
\be
Z_1 = \cube - {\pbth \over \cube},
\ee
and 
\be
Z_4 = {\sqrt{3}\over 2}\left[ {\pbth \over \cube} + \cube \right] 
\ee
respectively.
The quantities $p$ and $q$ are defined as
\be
p = (B_0 - {A^2_0\over 3}),
\ee
and
\be
q = ( 2 C_0 - {A_0 B_0\over 3} + {2\over 27} A^3_0 ).
\ee
The solution eqn.(\ref{sol}) for the average neutrino conversion 
probability ${\cal P}(t)$ is very complicated. So here we will consider
condition for the existence of the solution rather than going into 
the details of it. 
Since the probability is positive definite ($0\le {\cal P}(t)\le 1$),
the following condition,
\be 
-{q\over 2} + \sqrt{\left({p\over 3}\right)^3 + 
\left({q\over 2}\right)^2}~~ > ~~0,
\label{con1}
\ee
has to satisfy, 
otherwise $Z_1$ and $Z_4$ will be complex so also ${\cal P}(t)$. 
Then putting values of $p$ and $q$ in eqn.(\ref{con1}) we obtain
the condition
\be
4 H^2_{ax}(0) + H^2_{aa}(0) ~>~{4\over 3} (\ga^2 + \gaa^2 - \ga\gaa).
\label{con2}
\ee
Thus for the average neutrino conversion probability to be positive
definite, the above condition  has to satisfy, 
irrespective of the form of the neutrino 
potential and the magnetic field. 
Let us consider the neutrino propagation in the early universe hot 
plasma and the core of the
newly formed neutron star, where the above condition can be satisfied.

\section{Early Universe hot plasma}

In the early universe when the temperature $T~ >>~ 1$ MeV, 
all the particles are in thermal equilibrium\cite{koste}. 
In the early universe the magnetic flux lines are moving along
with the hot plasma. Because of the chaotic motion of the plasma, the
flux lines will be mixed up and twisted thus creating the randomness
in the magnetic field.
So the constant background field $B_0$ must be very small compared
to the random fluctuation term. Thus we can safely assume $B_0\simeq 0$ and
then $H_{ax}(0)\simeq 0$ and $\mu_{eff} B_{\parallel 0}\simeq 0$.
For small  particle-anti particle asymmetry in the early 
universe hot plasma the axial vector potential contribution can be  very small
as in a relativistic plasma the charged lepton/anti-lepton 
masses and their chemical
potentials are small compared to the kinetic energy term, thus the factor
$\mu_{eff}$ will also be  small\cite{raffelt1,valle}. So we can neglect the
longitudinal damping term for the magnetic field.
Then
eqn.(\ref{con2}) will be
\be
H_{aa}(0) ~>~ {2\over \sqrt{3}} \ga.
\label{eur}
\ee
The vector potential for a neutrino in the early universe 
hot plasma is\cite{notzold} 
\be
V_{vec}={\sqrt 2} G_F n_{\gamma}(T)\left [L - A{T^2\over M_W^2}\right ].
\ee
Here $n_{\gamma}(T)\simeq 0.244 T^3$ is the
photon number density and $A\simeq 55$. For temperature $T~>>~ m_e$  
the second non-local term
is greater than the first term for
very small particle-anti particle asymmetry and the second term is
\be
V_{vec}\simeq - 3.45\times 10^{-20}\left ({T\over MeV}\right )^5~ MeV,
\ee
for electron neutrino. 
For $\nu_{eL}\rightarrow\nu_{eR}$ process we assume that the right handed
neutrino produced is sterile and decouple from the system. 

Irrespective of its origin, the primordial magnetic field had a very large
value in the early universe. 
We assume that the primordial plasma consists of magnetic domain structure
with a size $L_0$ and the magnetic field is uniform and constant within
each domain and the field in different domains are randomly 
aligned\cite{raffelt1,semikoz}.
For homogeneous magnetic fields, the flux conservation indicates that
$B\propto T^2$. 
But detailed structure of random magnetic field profile in the 
early universe hot plasma will
depend on the complicated nature of the magneto-hydrodynamic 
equations\cite{branden}. 
For the root mean square field $B_{rms}=\sqrt{<B^2>}$, averaged over 
a volume $L^3~>>~L_0^3$ we assume a power law behavior\cite{vachas,enq},
\be
B_{rms}=B_n\left ({T\over T_0}\right )^2 \left ({L_0\over L}\right )^n,
\label{brms}
\ee
where $T_0$ is the temperature at some reference epoch and $B_n$ is the
corresponding field strength within a domain of size $L_0$. 
The maximal scale we have chosen is the horizon length
$L=l_H=M_{pl}/T^2$, where $M_{Pl}=1.22\times 10^{19}$ GeV is the Planck mass. 
We take here $B_n T_0^2 L_0^n$ to be constant.
If the primordial magnetic fields are to be the seed fields for the galactic
dynamo, then it should survive till the recombination epoch and this leads
to a minimal domain size\cite{cheng}, 
\be
L_0 \ge L_0^{min} \simeq 10^3 ~cm\times \left ( {MeV\over T}\right ).
\ee
Here we take $B_n=10^{24}$ Gauss and $T_0=T_{EW}=10^5$ MeV the electroweak
phase transition temperature\cite{semikoz}. 
Putting all these in eqn.(\ref{eur}) we obtain
\be
\mu ~\le ~ {1.15\times 10^{-15 + 12 n}\over (4.15)^n} \left ({T\over MeV}
\right )^{(1-n)} \left ({L_0\over cm}\right )^{-({1\over 2} + n)}.
\ee
The index parameter $n=0$ corresponds to the uniform magnetic field,
which is not physically very likely in the early universe hot plasma.
For the random magnetic field along the neutrino trajectory the index 
parameter would be $n=1/2$ and $n=3/2$ corresponds to 3-dimensional 
elementary  cells\cite{raffelt1,pas1,semikoz}. 
As the magnetic field is random in nature we take
$n=1/2$. For QCD phase transition temperature $T=T_{QCD}\simeq 200$ MeV we 
obtain $\mu\le 8\times 10^{-12}\mu_B$.

\section{Supernova Core}

Now let us consider the neutrino propagation in the supernova medium. 
As shown in eqn.(\ref{haa}) and  eqn.(\ref{hax}) we have
$H_{aa}(0) = (V_{vec}-\Delta + \mu_{eff}B_{\parallel 0})$ and
$H_{ax}(0) = \mu B_{\perp 0}$. Thomson and Dunkan\cite{thomson}
have shown that,
magnetic fields as strong as $10^{14}$ to $10^{16}$ Gauss might be generated
inside the core of the supernova due to a small scale dynamo mechanism. If
these fields are generated after core collapse, then it could be viewed
as random superposition of many small dipole of size $L_0\sim 1$ Km.
So we neglect here the $B_0$ term in the diagonal and non-diagonal parts.
Thus $H_{aa}(0) = (V_{vec}-\Delta)$ and $H_{ax}(0)=0$.
For $\nu_{eL}\rightarrow \nu_{eR}$ the vector potential 
experience by neutrino in the supernova medium is
\be
V_{vec} = 4\times 10^{-6} \rho_{14} f(Y_e)~ MeV,
\ee
where $f(Ye) = (3 Ye -1)$ is the electron
neutrino abundance factor and $\rho_{14}$ is the density in units of
$10^{14}~g/cm^3$. The right-handed electron neutrino being sterile
stream away from the supernova core. 
The $\Delta={5\times 10^{-15}/ E_{100}} 
\left ({{\Delta m^2}/ eV^2}\right )$ MeV, where the neutrino energy
$E_{100}$ is in units of 100 MeV. We can see that $\Delta ~<<~V_{vec}$
even for large $\Delta m^2$.
Inside the neutron star core, there are less number of electrons, so
the damping term $\gaa$ can be very small compared to the 
vector potential or the transverse damping term $\ga$. Assuming $\gaa$
to be very small inside the core we have $H_{aa}(0)\simeq V_{vec}$. 
Then eqn.(\ref{con2}) is reduced to the same condition as in the early
universe hot plasma
$ H_{aa}(0) ~>~2\ga/\sqrt{3}$. 
Inside the core the damping term is
\be
\ga = 2.2\times 10^{-18} {\mu_0}^2 \left ({B^2_{rms}\over G^2}\right )
\left ({L_0\over cm}\right )~MeV,
\ee
where ${\mu_0}$ is in units of $\mu_B$ the Bohr magneton. Putting the
values of $H_{aa}(0)$ and $\ga$ in eqn.(\ref{eur}) we obtain
for the magnetic moment
\be
\mu_0 ~\le ~{{1.25\times 10^6 \left (\rho_{14} |f(Y_e)|\right )^{1/2}}\over 
\left ({B_{rms}\over G}\right ) \left ({\L_0\over cm}\right )}.
\ee
For $B_{rms}\sim 10^{16}$ Gauss, $L_0\sim 1$ Km, $\rho_{14}\sim 8$ 
and the electron abundance 
factor $Y_e\sim 0.3$ inside the neutron star\cite{pas2}, 
the magnetic moment comes
out to be $\mu ~\le 3.5\times 10^{-13}\mu_B$.
For smaller value of the magnetic field, the magnetic moment will be larger.

\section{Discussion}

Assuming the magnetic field has a random fluctuation over the mean
value, we have derived the average probability equation for 
neutrino conversion/spin precession in the magnetized plasma. 
As a consequence of the random fluctuation in the 
magnetic field the transverse and longitudinal magnetic field damping
are getting mixed up. We have assume a delta correlation for 
the random magnetic field domains and
obtain the solution for the probability equation.
The definiteness of the probability is invalid if the condition in 
eqn.(\ref{con2}) is not satisfied. 
We assume the magnetic field to be purely random in nature, 
in the early universe hot plasma as well as 
in supernova medium (inside the newly born neutron star core) and consider
the process $\nu_{eL}\rightarrow\nu_{eR}$.
At the QCD phase transition temperature $T\simeq 200$ MeV we obtain the
neutrino magnetic moment $\mu\le 8\times 10^{-12}\mu_B$. Inside the supernova
core we consider the same process $\nu_{eL}\rightarrow\nu_{eR}$ and for 
$B_{rms}\simeq 10^{16}$ Gauss, obtain $\mu ~\le 3.5\times 10^{-13}\mu_B$.
In this estimate we have neglected the contribution from the
axial vector potential as well as the contribution due to the constant
magnetic field in the plasma. We have also neglected the contribution from 
the longitudinal damping term. So inclusion of all these terms
might improve the  constraints on the magnetic moment. Apart from that
the magnetic field profiles for the supernova and early universe are
very much speculative, so the upper limit might change for other magnetic 
field profiles.

       I am thankful to Prof. V. B. Semikoz and Prof. J. W. F. Valle for 
many helpful discussions during the initial stage of the work. 


\end{document}